# AI assisted optimization of integrated waveguide polarizers containing 2D reduced graphene oxide

Rong Wang,[1,2,†] Yijun Wang,[1,†] Di Jin,[2,3,*] Junkai Hu,[1,2,3] Wenbo Liu,[4,5] Yuning Zhang,[6] Duan Huang,[1,*] Jiayang Wu,[2,3] Baohua Jia,[3,4,5,*] and David J. Moss[2,3,*]

[1] School of Automation, Central South University, Changsha 410083, China

[2] Optical Sciences Centre, Swinburne University of Technology, Hawthorn Victoria 3122, Australia

[3] Australian Research Council (ARC) Centre of Excellence in Optical Microcombs for Breakthrough Science (COMBS)

[4] School of Science, RMIT University, Melbourne, Victoria 3000, Australia

[5] Australian Research Council (ARC) Industrial Transformation Training Centre in Surface Engineering for Advanced Materials (SEAM), RMIT University, Melbourne, Victoria 3000, Australia

[6] School of Physics, Peking University, Beijing, 100871, China

[†] These authors contributed equally.

*dijin@swin.edu.au, duanhuang@csu.edu.cn, baohua.jia@rmit.edu.au, dmoss@swin.edu.au

**Abstract:** Reduced graphene oxide (rGO) exhibits strong anisotropic light absorption and high compatibility with photonic integrated chips, making it a promising material for implementing high-performance on-chip polarization-selective devices. The performance of rGO integrated waveguide polarizers is highly dependent on the waveguide geometry, and achieving optimal performance requires exploring a large parameter space, making conventional mode simulation methods computationally demanding. Here, we propose and demonstrate a machine learning framework based on fully connected neural networks (FCNNs) to map the dependence of the polarizer figure of merit (FOM) on the waveguide geometry. Once trained by using a small dataset of low-resolution mode simulation results, the FCNN framework can rapidly and accurately predict FOM values across a large structural parameter space with high resolution. Results show that this method can reduce overall computing time by more than 4 orders of magnitude as compared to the mode simulation methods, and achieve high prediction accuracy with an average deviation (*AD*) below 0.05. These results highlight the FCNN-based machine learning framework as an efficient tool for the design and optimization of rGO integrated waveguide polarizers.

## 1. Introduction

Polarization control is fundamental to optical technologies, and polarizers serve as essential components that enable this function by allowing one polarization state to propagate and suppress the orthogonal state [1-4]. Recently, on-chip integration of two-dimensional (2D) materials with strong anisotropic light absorption and broadband optical response has emerged as an attractive approach for realizing integrated polarizers with wide operation bandwidth and high polarization selectivity [3-7]. Particularly, reduced graphene oxide (rGO) has shown several unique advantages among various 2D materials [8-11]. First, it exhibits much stronger anisotropic light absorption than graphene oxide (GO), enabling enhanced polarization selectivity for rGO polarizers [10, 12, 13]. Second, rGO can be easily produced by reducing GO, retaining the benefits of solution-based and transfer-free methods for on-chip integration of 2D GO films and offering excellent compatibility with photonic integrated chips [14-16]. Finally, in contrast to GO that typically undergoes reduction under elevated temperatures and high optical powers, rGO exhibits significantly improved thermal stability and power endurance, making it particularly advantageous for high-power applications [8, 11, 16].

The performance of rGO integrated waveguide polarizers, quantitatively evaluated by the polarizer figure of merit (FOM), is highly dependent on the waveguide structural parameters [8]. In conventional methods, optimizing the structural parameters to achieve high FOM values requires extensive mode simulations across large structural parameter spaces. Such simulations rely on numerical mode simulation software and are computationally demanding, especially since accurate simulations of waveguides incorporating 2D materials require extremely fine meshing [17-19].

Recently, rapid advances in artificial intelligence (AI) technology are revolutionizing the modeling and design of optical devices [20-23]. Unlike conventional simulation approaches that depend on iteratively solving Maxwell's equations, AI significantly enhances computational efficiency by constructing neural networks that directly map structural parameters to optical responses, thereby capturing underlying physical relationships [24-27]. This transformative paradigm offers broad applicability, demonstrating notable strengths in tackling complex design challenges and optimizing sophisticated photonic structures [21, 23, 28-30]. Up to now, AI-driven methods have achieved considerable success in designing diverse functional devices, including metasurfaces [24, 31-34], nonlinear optical devices [35, 36], electro-optic modulators [37], photodetectors [38-40], and quantum optical devices [41-43].

In this work, we propose a machine learning framework based on fully connected neural networks (FCNNs) to optimize the performance of rGO integrated waveguide polarizers. Trained by using a small dataset of low-resolution mode simulation results, the FCNN framework can predict the FOM values across a large structural parameter space with high resolution. Compared with traditional mode simulation methods, this approach not only reduces overall computing time by more than 4 orders of magnitude, but also achieves high prediction accuracy with an average deviation ($AD$) below 0.05. In addition, we construct nine training datasets with different sizes, and analyze the trade-off between improvements in prediction accuracy and the cost of building the training dataset. These results verify the effectiveness of the FCNN-based machine learning framework in efficiently designing and optimizing rGO integrated waveguide polarizers.

## 2. Device structure

Fig. 1a illustrates the schematic of an integrated waveguide polarizer based on a silicon photonic waveguide coated with a monolayer rGO film. The cross section of this hybrid waveguide is shown in Fig. 1b, where $W$ and $H$ represent the width and height of the silicon waveguide, respectively, and $n$, $k$, and $d$ represent the refractive index, extinction coefficient, and thickness of the rGO film, respectively. These parameters play a critical role in determining the performance of the rGO integrated waveguide polarizers.

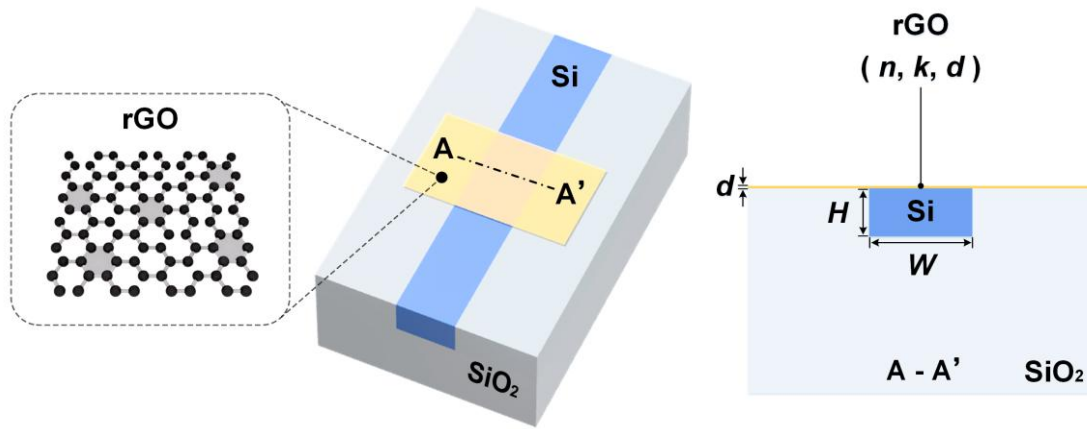

**Fig. 1. a,** Schematic illustration of an integrated waveguide polarizer consisting of a silicon (Si) photonic waveguide coated with a monolayer reduced graphene oxide (rGO) film. **b,** The cross-sectional view of hybrid waveguide is shown in the right panel, where $W$ and $H$ denote the width and height of the silicon waveguide, respectively, and $n$, $k$, and $d$ represent the refractive index, extinction coefficient, and thickness of the rGO film, respectively.

rGO [44-48] can be derived from GO [4, 6, 49-51] through reduction processes that remove oxygen-containing functional groups (OFGs) such as hydroxyl, epoxide, carbonyl, and carboxylic groups [52, 53]. In contrast to graphene, which has a low solubility, GO containing hydrophilic OFGs can be readily dispersed in water and processed in solution [14, 54]. Recently, a solution-based, transfer-free method has been developed for on-chip integration of 2D GO films [12, 13, 54, 55], which allows layer-by-layer film coating in large areas with precise control of the film thickness. The GO coated on integrated devices can be easily reduced to form rGO by using various methods, such as thermal reduction, chemical reduction, laser reduction, and microwave reduction [15]. During the reduction process, the restoration of $sp^2$-hybridized carbon domains results in a decreased optical bandgap and changes in the material properties such as optical absorption and refractive index [8, 11]. Different degrees of rGO reduction are achieved depending on the extent to which OFGs are removed. When the OFGs are completely removed and no residual groups remain on the carbon network, the bandgap approaches zero, and the material properties closely resemble those of graphene [52, 56]. Therefore, the reduction of GO into rGO also provides an appealing route for mass-producing graphene-like materials through solution-based processing [57, 58].

The hybrid waveguide shown in Fig. 1a exhibits significantly stronger light absorption for transverse electric (TE, in-plane) polarization than for transverse magnetic (TM, out-of-plane) polarization. This behavior arises from the interaction between the evanescent field of the silicon waveguide and the 2D material film, which possesses strong anisotropy in its light absorption [8, 10, 11]. In addition, owing to the broadband anisotropic absorption of rGO spanning the visible to infrared region, the hybrid waveguide exhibits a substantially wider operational bandwidth in contrast to bulk-material-based optical polarizers, which are generally limited to bandwidths below 100 nm [3, 7]. Although rGO inherently exhibits strong polarization selectivity, the overall polarizer's performance is strongly governed by the waveguide geometry, as structural parameters such as waveguide width and height directly determine the modal field distribution and the strength of the light-matter interaction. In the following, we will use an FCNN-based machine learning framework to optimize the performance of rGO integrated waveguide polarizers.

**3. FCNN-based machine learning framework**

Fig. 2 illustrates the process flow for using a machine learning approach to predict FOM of an integrated waveguide polarizer coated with a monolayer rGO film. In this approach, mode simulations performed with low-resolution structural parameters ($W$, $H$) are used to train a predictive framework, which is then applied to predict the polarizer performance across a high-resolution design space ($W'$, $H'$), thereby enabling efficient exploration of the structural parameter space and facilitating the identification of device geometries with high performance. This process consists of three steps.

First, the optical parameters of the rGO film ($n$, $k$, $d$) are experimentally measured for use in subsequent mode simulations. For example, rGO film thickness can be determined using atomic force microscopy (AFM) [11]. The TE- and TM-polarized refractive indices ($n_{TE}$, $n_{TM}$) and extinction coefficients ($k_{TE}$, $k_{TM}$) of the rGO film can be extracted by fitting the transmission spectra of rGO integrated microring resonators (MRRs) [9, 59] using the scattering matrix method [60, 61].

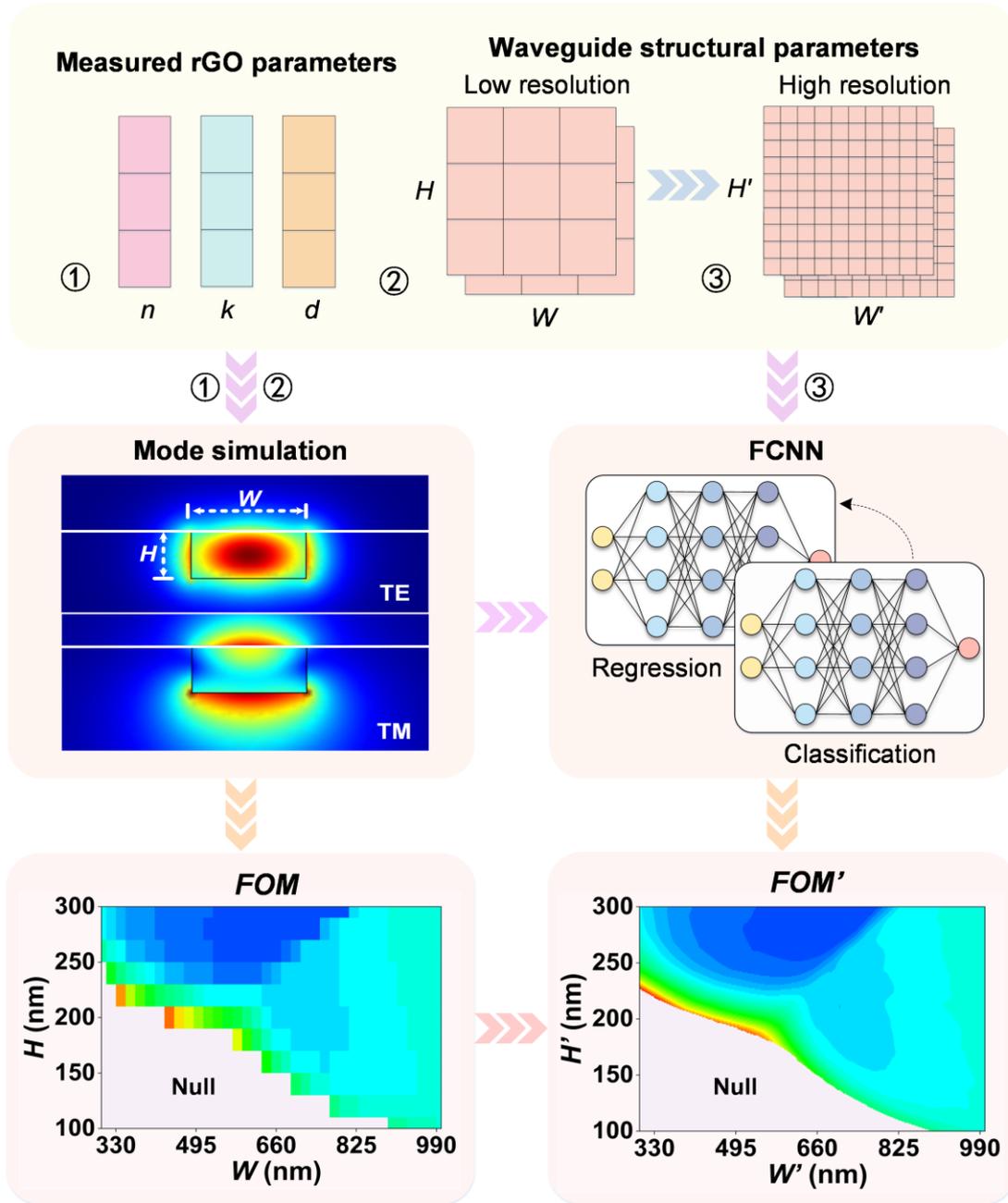

**Fig. 2.** Schematic illustration of a fully connected neural network (FCNN) framework, which maps low-resolution structural parameters ($W$, $H$) obtained from mode simulations to high-resolution parameters ($W'$, $H'$) for predicting the polarizer figures of merit ($FOM'$).

Second, using the measured rGO parameters, mode simulations are conducted for the hybrid waveguides with low-resolution ($W$, $H$) sets. For small $W$ or $H$ values where the corresponding TE or TM modes fail to converge, indicating that these dimensions meet the mode cut-off condition and the modes cannot physically exist. In such sets, the corresponding (*W*, *H*) is recorded as 'Null'. For converged modes, the power propagation losses (dB/cm) of the hybrid waveguide can be calculated by [9, 11, 59]

$$PL_{TE} = -10 \cdot \log_{10}\left\{\left[exp(-2\pi \cdot k_{TE,\,eff} \cdot L / \lambda)\right]^2\right\} \quad (1)$$

$$PL_{TM} = -10 \cdot \log_{10}\left\{\left[exp(-2\pi \cdot k_{TM,\,eff} \cdot L / \lambda)\right]^2\right\} \quad (2)$$

where $k_{TE,\,eff}$ and $k_{TM,\,eff}$ are the imaginary parts of effective refractive indices for the TE and TM modes, respectively, *L* = 1 cm is the waveguide length, and *λ* is the wavelength of light. The polarizer figure of merit (*FOM*) is then calculated by [6]

$$FOM = PDL / EIL = (PL_{TE} - PL_{TM}) / PL_{TM} \quad (3)$$

where *PDL* is the power dependent loss, defined as the difference in insertion loss between the TE and TM modes. This metric has been widely applied to quantify the polarization selectivity of optical polarizers [4, 8, 49]. *EIL* represents the minimum excess insertion loss induced by the rGO film, equivalent to the excess insertion loss for TM polarization. In 2D-material-based optical polarizers, the 2D material films provide high polarization selectivity as well as introduce excess insertion loss. Therefore, the *FOM* defined in Eq. (3), which reflects the trade-off between these two factors, is commonly used to evaluate the performance of 2D-material-based optical polarizers [4, 6, 7].

Finally, the recorded 'Null' for (*W*, *H*) corresponding to non-converged modes and the calculated *FOM* values for (*W*, *H*) corresponding to converged modes are used to construct the training dataset for the FCNN framework. The modelling involves a low-dimensional parametric classification and regression problem, for which an FCNN framework is adopted, rather than more complex architectures developed for image- or sequence-based problems (*e.g.,* convolutional neural networks or recurrent neural networks). The framework is designed with two subnetworks, one responsible for mode-convergence classification and the other for FOM prediction. After training on low-resolution structural parameters ($W$, $H$), the framework can determine whether the TE or TM modes converge and, for converged cases, predict the corresponding *FOM*'s across high-resolution ($W'$, $H'$) structural parameter sets. In this way, the FCNN framework enables efficient exploration of the full design space and provides a predictive foundation for identifying waveguide geometries with high performance. The details of the FCNN framework will be presented in Fig. 3. For clarity in comparison, the following discussion employs the same yet slightly different manner to label the training and test datasets parameters. For instance, *W*, *H*, and *FOM* refer to the parameters for the training dataset, whereas *W'*, *H'*, and *FOM'* correspond to those for the test dataset.

To highlight the necessity of the proposed machine-learning-assisted approach, it is instructive to first examine the limitations of conventional methods. Conventional methods rely on commercial mode simulation software (such as COMSOL Multiphysics and Lumerical FDTD) to exhaustively evaluate all combinations of *W* and *H*. For example, scanning over $W \in [300, 1000]$ nm and $H \in [100, 300]$ nm at 1-nm resolution would require over 140,000 individual simulations, with each simulation typically taking 5–7 minutes. This results in prohibitively high computing time, particularly because the ultra-thin rGO films (with thicknesses typically on the order of 1 nm) require ultra-fine mesh resolution to ensure simulation accuracy.

By contrast, the proposed machine-learning framework extracts and models the relationships between structural parameters and modal behavior from a small number of (*W*, *H*) sets. Once trained, the framework can rapidly predict *FOM'* for an arbitrary high-resolution (*W'*, *H'*). Furthermore, sweeping the entire high-resolution (*W'*, *H'*) using our method adds minimal extra time compared to predicting a single set. For instance, predicting the *FOM'* for one set with converged mode takes 40–80 ms, whereas sweeping over 140,000 sets of (*W'*, *H'*) takes only 25–35 seconds, with each additional prediction contributing less than 1 ms to the total computing time. In addition to significantly saving computing time, our method also achieves high prediction accuracy. For example, using 396 sets of (*W*, *H*) as the training dataset, our method can achieve a high accuracy of ~99.0% in predicting mode convergence and a low average deviation (*AD*) of 0.043 across *FOM'* predictions for more than 140,000 (*W'*, *H'*) sets.

In this work, we develop a computer-aided design (CAD) approach for rGO integrated waveguide polarizers using a commonly adopted FCNN framework in ML. The material parameters employed in the simulations, including *n*, *k*, and *d*, are obtained from experimental measurements [8, 11, 60], ensuring that the predicted high-*FOM* regions are physically realistic and provide guidance for practical device design. Unlike ML-based inverse design approaches that attempt to directly learn mappings from target performance to device structure [33, 62, 63], the proposed method follows a forward design strategy that first identifies physically meaningful regions of the global parameter space and then learns the relationship between structural parameters and device performance within this constrained space. Since the forward mapping from structure to performance is physically single-valued, this strategy naturally avoids the non-uniqueness [64-66] associated with inverse design of optical devices. In addition, by restricting the learning task to a physically consistent and lower-complexity parameter subspace, the model can capture global physical trends with higher sample efficiency, thereby significantly mitigating the reliance on large-scale simulation datasets.

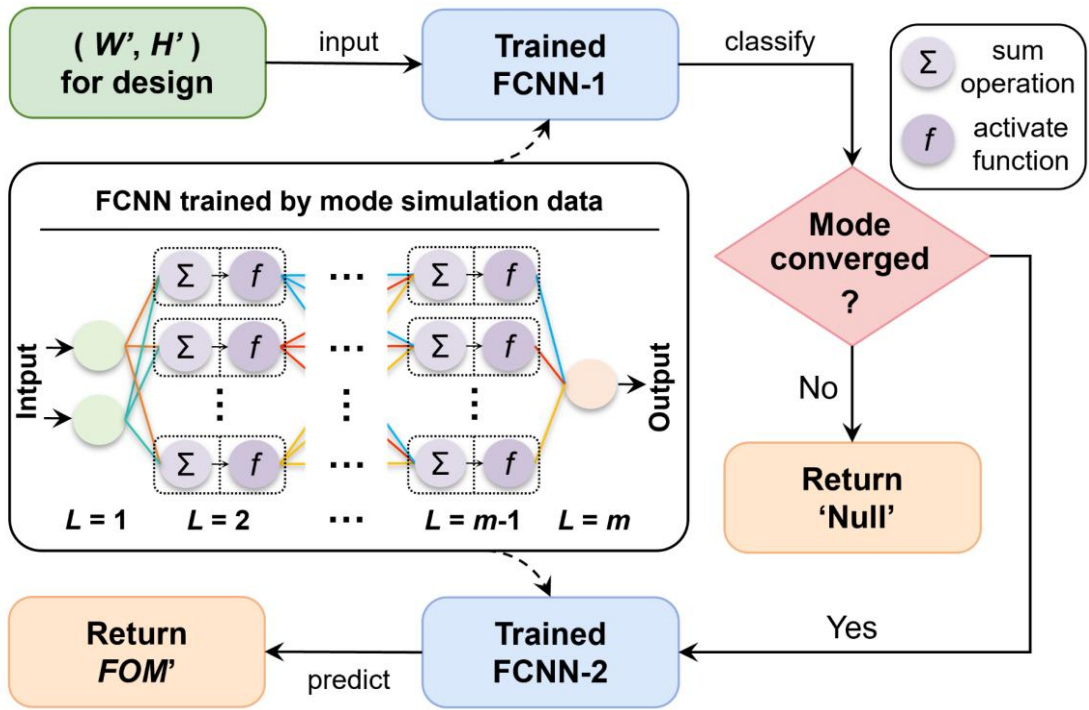

**Fig. 3.** Schematic illustration of the FCNN framework, which comprises two subnetworks FCNN-1 and FCNN-2 for identifying mode convergence and predicting polarizer *FOM*'s, respectively. Inset in the middle panel illustrates the architecture of either subnetwork, consisting of an input layer ($L = 1$), multiple hidden layers ($L = 2, 3, …, m$-1), and an output layer ($L = m$).

Fig. 3 illustrates the FCNN framework used to predict the *FOM'* of rGO integrated waveguide polarizers over a high-resolution structural parameter space. The framework processes the input high-resolution test dataset (*W'*, *H'*) in two sequential stages. In the first stage, FCNN-1 classifies whether the TE and TM modes are converged. Parameter sets for which either the TE or TM mode fails to converge, indicating that the corresponding guided mode cannot be sustained under those dimensions, are classified as 'Null'. In the second stage, only those (*W*', *H*') combinations identified as mode-converged are fed into FCNN-2, which predicts the corresponding polarizer *FOM*'. This two-step structure enables rapid screening of the design space by eliminating invalid configurations before performance evaluation.

Both FCNN-1 and FCNN-2 employ a similar architecture consisting of an input layer, three hidden layers, and an output layer, as illustrated in the middle inset of Fig. 3. The input layer includes two neurons and receives the waveguide structural parameters (*W*, *H*). FCNN-1 consists of three hidden layers with 32, 64, and 32 neurons, whereas FCNN-2 consists of three hidden layers with 64, 128, and 64 neurons. Neurons in each layer are fully connected to neurons in the next layer through weighted links. Each neuron computes a weighted summation of inputs from the previous layer, adds a bias term, and applies an activation function to generate its output. The output layer of each subnetwork consists of a single neuron but serves a task-specific role. Specifically, FCNN-1 produces a binary classification probability for mode convergence identification, whereas FCNN-2 outputs a regression prediction of the *FOM*'. This architecture is standard for FCNNs, similar to that in Refs [67, 68], and is suitable for modeling complex mapping between structural parameters and device performance.

Before performing the FCNN framework in Fig. 3 to identify converged modes and predict their corresponding polarizer *FOM*' for high-resolution (*W*', *H*') sets, both subnetworks were trained using a low-resolution training dataset (*W*, *H*). The training began with FCNN-1 for mode convergence identification. A training dataset consisting of ($W$, $H$, $C$) pairs was constructed, with $C$ encoded as 1 for converged and 0 for non-converged modes. This dataset was then normalized and fed to the input layer, generating the input vector $h_1 = (W^{norm}, H^{norm})$. For a given layer $p$, the output $h_{p,j}$ of the $j$-th neuron is computed as [69-71]

$$h_{p,j} = f\left(\sum_{i=1}^{q} w_{p,ij} \cdot h_{p-1,i} + b_{p,j}\right) \tag{4}$$

where $q$ is the total number of neurons in the ($p$-1)-th layer, $h_{p-1,i}$ is the output of the $i$-th neuron in the layer $p$-1, $w_{p,ij}$ is the connection weight from the $i$-th neuron in the ($p$-1)-th layer to the $j$-th neuron in the $p$-th layer, $b_{p,j}$ is the bias of the $j$-th neuron in the current layer, and $f$ is the activation function. ReLU [72] activation function is applied to all hidden layers, but a Sigmoid activation function is used in the output layer to generate the convergence probability $y \in [0, 1]$. When $y > 0.5$, the input ($W$, $H$) is classified as a converged mode, if not, it is classified as non-converged. During the training, we first randomly split the dataset into training and validation sets in an 8:2 ratio and employed the Adam optimizer [73] with a learning rate of $1 \times 10^{-4}$ and a batch size of 128. To avoid overfitting, we adopted multiple strategies, including employing an early stopping criterion of 50 epochs to prevent excessive training, applying training data noise regularization to improve robustness, and continuous monitoring of the loss curves to ensure stable convergence. FCNN-1 evaluated its results against the ground-truth labels using the binary cross-entropy (BCE) loss and accuracy [25, 74]. The gradients were backpropagated to iteratively update the weights $w$ and biases $b$ across all layers, allowing the network to learn the underlying input-output mapping.

Following the completion of FCNN-1 training, FCNN-2 was trained for polarizer *FOM'* prediction using only mode-converged (*W*, *H*, *FOM*) sets. FCNN-2 adopts the same fully connected structure as FCNN-1 but uses a ReLU activation function in the output layer to predict continuous *FOM'* values. The model was also trained using the Adam optimizer [73], with the dataset divided into training and validation sets following an 8:2 ratio and employing the same learning rate and overfitting mitigation strategies as those adopted for FCNN-1, but with a different batch size of 16. Training was performed via backpropagation, and the prediction accuracy was evaluated using regression metrics, including root mean square error (RMSE) and $R^2$-score [25, 74, 75].

## 4. Results and discussion

Based on the FCNN framework in Fig. 3, we trained FCNNs for optimizing the *FOM* for rGO integrated waveguide polarizers across varying waveguide width *W* and height *H*. To investigate the influence of training data density on prediction accuracy, three training datasets were constructed using uniform step sizes of Δ = 80 nm, 40 nm, and 20 nm for both *W* and *H* between adjacent parameters within the ranges of *W* ∈ [300, 1000] nm and *H* ∈ [100, 300] nm. A smaller step size Δ corresponds to a larger number of (*W*, *H*) sets in the training dataset. For example, for mode-convergence identification, the dataset with Δ = 20 nm contains 396 sets, whereas the dataset with Δ = 80 nm only has 27 sets. The ranges of *W* and *H* were chosen to approximately span the convergence boundaries of the fundamental TE and TM modes for single-mode integrated waveguides operating near 1550 nm. In addition, a high-resolution test dataset, generated with Δ' = 1 nm within the same ranges of *W'* ∈ [300, 1000] nm and *H'* ∈

[100, 300] nm, was used to evaluate the trained FCNNs for both mode-convergence identification and polarizer *FOM'* prediction.

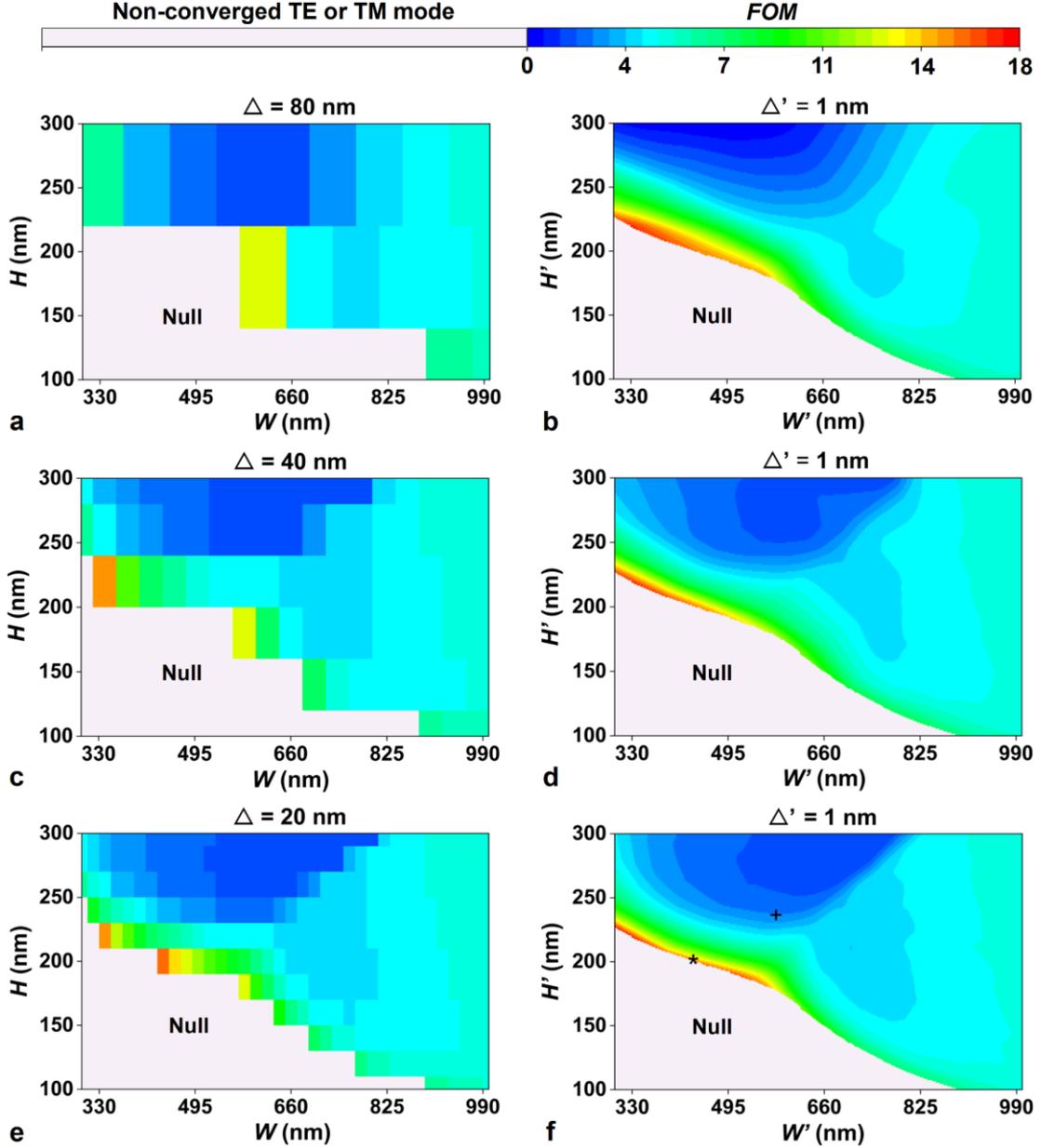

**Fig. 4. a, c, e,** *FOM* of rGO integrated waveguide polarizer versus low-resolution (*W*, *H*) with Δ = 80 nm, 40 nm, and 20 nm, respectively, where Δ is the step size between adjacent waveguide parameters within ranges of *W* ∈ [300, 1000] nm and *H* ∈ [100, 300] nm. The *FOM* values were calculated based on mode simulations and the 'Null' regions denote the cases of non-converged TE or TM modes in the simulations. **b, d, f,** *FOM'* versus high-resolution (*W'*, *H'*) with Δ' = 1 nm. The *FOM'* values were predicted using FCNN-2, which was trained with the data in **a, c, e,** respectively. Regions labeled 'Null' correspond to non-converged TE or TM modes, as identified by FCNN-1.

Fig. 4a shows the *FOM* of rGO integrated waveguide polarizer versus low-resolution (*W*, *H*) with a step size of Δ = 80 nm. The *FOM* values were calculated based on mode simulations using commercial software (COMSOL Multiphysics), and the 'Null' region corresponds to the (*W*, *H*) with non-converged TE or TM modes in the simulations. The corresponding GO film parameters (*n*, *k*, *d*) used in our mode simulations were obtained from the experimental measurements reported in Ref. [8], where the rGO film was thermally reduced from GO with a high reduction degree. In that work, the silicon waveguide was first coated with a monolayer GO film and subsequently heated on a hotplate at 150 °C for 15 min. The thickness (*d*) of the rGO film was set to 1 nm. The refractive index (*n*) and extinction coefficient (*k*) of the rGO film at 1550 nm were $n_{TE}$ = ~2.1 and $k_{TE}$ = ~0.194 for TE polarization, and $n_{TM}$ = ~1.97 and $k_{TM}$ = ~0.0272 for TM polarization. For Δ = 80 nm, the (*W*, *H*) parameter space contains only 27 sets. As a result, Fig. 4a appears as large, discretized patches, reflecting the limited resolution of the training dataset.

Fig. 4b shows *FOM'* versus high-resolution (*W'*, *H'*) with Δ' = 1 nm, obtained using the FCNN framework in Fig. 3, which was trained on the low-resolution dataset shown in Fig. 4a. The 'Null' region indicates cases of non-converged TE or TM modes identified by FCNN-1, whereas the *FOM'* values in the convergent region were predicted by FCNN-2. Compared with Fig. 4a, the results in Fig. 4b exhibit much higher resolution, where the convergence boundaries and the trend of *FOM'* variation remain consistent with the low-resolution results obtained from mode simulations. This demonstrates that the trained FCNN framework can effectively predict high-resolution

*FOM'* by capturing the dependence of mode convergence and *FOM* on the waveguide structural parameters. We also note that the change of *FOM'* with either *W'* or *H'* is non-monotonic, mainly resulting from non-monotonic changes in the mode overlap with the rGO film. This complex dependence cannot be accurately modeled by simple interpolation or polynomial fitting, highlighting the challenges in optimizing the performance of such devices. In contrast, our machine learning approach can efficiently capture complex dependencies from a low-resolution dataset, providing a powerful tool for optimizing device structural parameters to achieve maximum *FOM'*.

Figs. 4c and 4e show the *FOM* of rGO integrated waveguide polarizer versus (*W*, *H*) with training step sizes of $\Delta$ = 40 nm and 20 nm, respectively. Figs. 4d and 4f show the corresponding *FOM'* versus high-resolution test (*W'*, *H'*) with $\Delta'$ = 1 nm. For $\Delta$ = 40 nm and 20 nm, the training datasets contain 108 and 396 sets of (*W*, *H*), respectively. Compared with Fig. 4a, these larger training datasets provide more detailed mode simulation results for training our FCNN framework. Similar to Fig. 4b, the higher-resolution results in Figs. 4d and 4f also capture the variation trends of the convergence boundaries and *FOM* obtained from mode simulations. This confirms that the FCNN framework can reliably infer the high-resolution performance of rGO integrated waveguide polarizer from low-resolution training data, thereby enabling efficient and accurate exploration of the design parameter space.

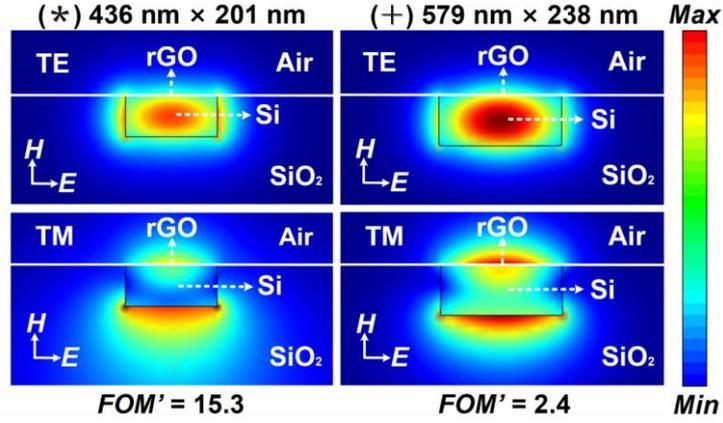

**Fig. 5.** TE and TM mode profiles corresponding to a high- and a low-*FOM'* point in **Fig. 4f**, marked by '✶' and '+', respectively.

Fig. 5 shows the TE and TM mode profiles corresponding to a high- and a low-*FOM'* point in Fig. 4f marked by '✶' and '+', respectively. The corresponding waveguide structural parameters (*W'*, *H'*) are (436 nm, 201 nm) and (579 nm, 238 nm), with *FOM'* values of ~15.3 and ~2.4, respectively. Mode simulations show *FOM* values of ~15.4 and ~2.4 for the same structural parameters, demonstrating good agreement with the predictions and confirming the accuracy of the FCNN framework in capturing polarizer's performance. We also note that the highest *FOM'* values in Fig. 4f are achieved near the mode convergence boundary, for example, the *FOM'* value at (439 nm, 200 nm) is ~15.7. Although selecting (*W*, *H*) exactly at the convergence boundary can maximize the polarizer *FOM*, this configuration typically results in a narrow operational bandwidth due to the strong wavelength dependence of mode cutoff. For practical polarizers, the trade-off between an increased polarizer *FOM* and a decreased operation bandwidth should be balanced. A practical solution is to choose (*W*, *H*) slightly offset from the convergence boundary, which can provide a relatively high *FOM* together with a minor decrease in the operation bandwidth. These results further reflect the complexity in optimizing the performance of 2D-material-based optical polarizers.

Table 1. Various training datasets ($W$, $H$) for the FCNN-based framework.

| Dataset No. | $\Delta H$ (nm)[a] | $\Delta W$ (nm)[a] | Number of training sets for FCNN-1 | Number of training sets for FCNN-2 |
|---|---|---|---|---|
| 1 | 80 | 80 | 27 | 17 |
| 2 | 40 | 80 | 54 | 39 |
| 3 | 20 | 80 | 99 | 73 |
| 4 | 80 | 40 | 54 | 32 |
| 5 | 40 | 40 | 108 | 75 |
| 6 | 20 | 40 | 198 | 141 |
| 7 | 80 | 20 | 108 | 64 |
| 8 | 40 | 20 | 216 | 150 |
| 9 | 20 | 20 | 396 | 281 |

[a]$\Delta W$ and $\Delta H$ are the step sizes used to uniformly sample $W \in [300, 1000]$ nm and $H \in [100, 300]$ nm, respectively, for establishing the training dataset ($W$, $H$).

Given that the mode simulation results are still needed for establishing the training dataset for the FCNN-based framework, the size of the dataset becomes a critical factor influencing both the prediction accuracy and the cost of data preparation. In principle, a larger training dataset with a finer resolution of the waveguide structural parameters enables higher prediction accuracy, but it also requires more time for mode simulations to generate the training dataset, resulting in a trade-off between them. In the following, we analyze this trade-off by comparing nine training datasets of ($W$, $H$) with different sizes. Table 1 summarizes the step size combinations ($\Delta W$, $\Delta H$) for these datasets designated as Nos. 1 – 9, each obtained by uniformly sampling $W \in [300, 1000]$ nm and $H \in [100, 300]$ nm. Decreasing either $\Delta W$ or $\Delta H$ increases the number of training samples, thereby enabling a quantitative assessment of how selecting step size $\Delta W$ and

Δ*H* influences the performance of the FCNN framework. When the step size is 80 nm for both Δ*W* and Δ*H* (*i.e.*, dataset No. 1), the numbers of training (*W*, *H*) sets for FCNN-1 and FCNN-2 are 27 and 17, respectively. As the step size decreases, the number of training (*W*, *H*) sets increases. When step size decreases to 20 nm for both Δ*W* and Δ*H* (*i.e.,* dataset No. 9), the training (*W*, *H*) sets for FCNN-1 and FCNN-2 increase to 396 and 281, respectively.

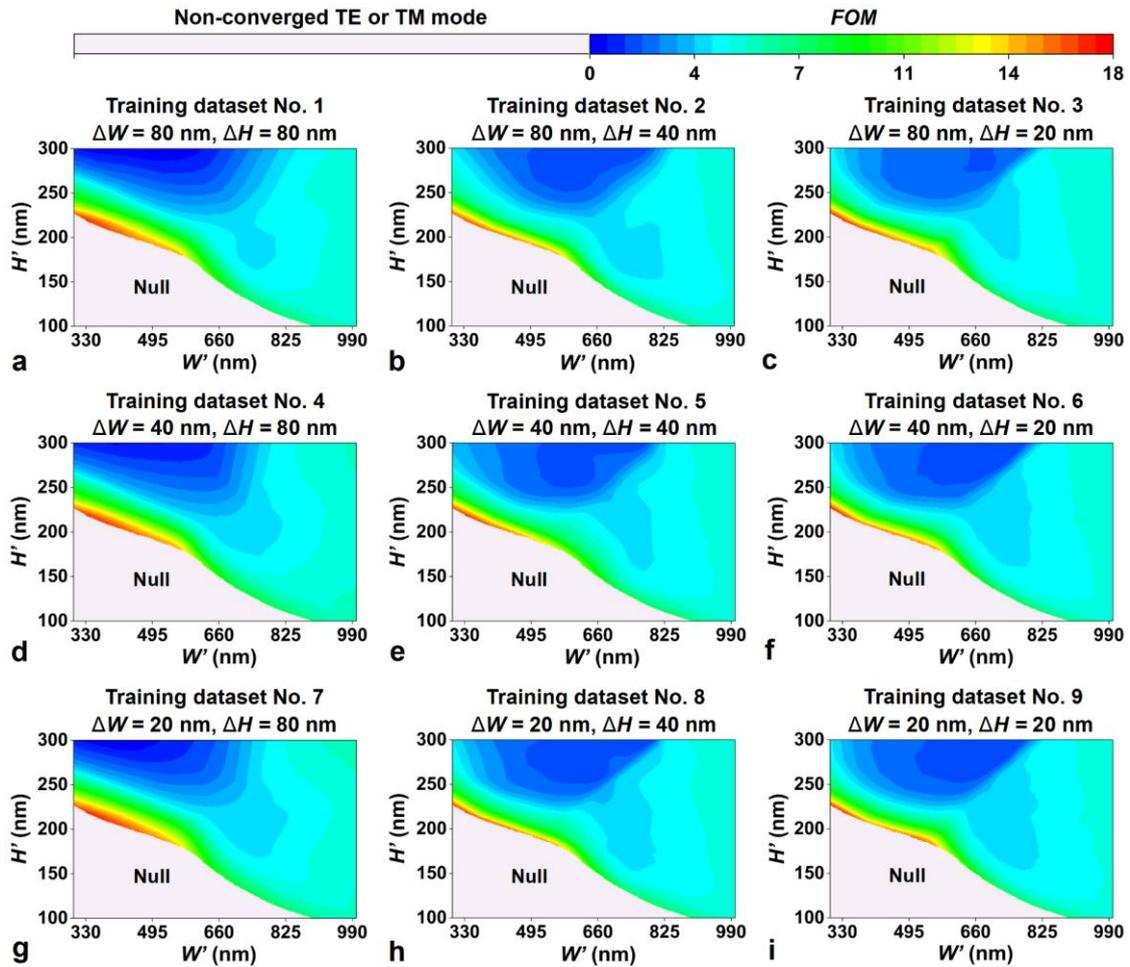

**Fig. 6. a-i,** *FOM'* versus high-resolution (*W'*, *H'*) with Δ' = 1 nm. The *FOM'* values were predicted using FCNN-2, which was trained using the datasets No. 1 – No. 9 listed in **Table 1,** respectively. The 'Null' regions denote the cases of non-converged TE or TM modes, which were predicted using FCNN-1.

Fig. 6 presents the predicted *FOM'* distributions obtained from the nine training datasets listed in Table 1. These results are evaluated using the same high-resolution test space with a step size of Δ' = 1 nm. Each subfigure corresponds to a specific (ΔW, ΔH) combination. The 'Null' region indicates cases of non-converged TE or TM modes identified by FCNN-1, whereas the *FOM*' values in the convergent region were predicted by FCNN-2.

In Figs. 6a–6c, ΔW is fixed at 80 nm, whereas ΔH decreases from 80 nm to 20 nm. Although the overall position of the mode-convergence boundary remains nearly unchanged, the predicted *FOM*' distribution becomes progressively smoother with more continuous gradient transitions as ΔH is reduced. This indicates that reducing the step size ΔH improves the granularity of the *FOM*' distribution without significantly affecting the convergence boundary. A similar trend is observed in Figs. 6a, 6d, and 6g, where ΔH is fixed at 80 nm but ΔW decreases from 80 nm to 20 nm. The convergence boundary maintains a consistent shape, whereas the internal *FOM*' distribution grows more detailed with decreasing ΔW. When both ΔW and ΔH are reduced simultaneously (Figs. 6a, 6e, and 6i), the *FOM*' distribution achieves the highest level of smoothness and detail, clearly delineating high-performance regions. This confirms that fine selection in both dimensions enhances the FCNN framework's prediction accuracy and enables more accurate identification of optimal structural parameters.

In Fig. 7, we compare the accuracy of FCNN-1 and the average deviation (*AD*) of FCNN-2 for various combinations of ΔH and ΔW corresponding to the training datasets in Fig. 6. The accuracy of FCNN-1 is defined as the proportion of correctly classified cases for mode convergence identification, whereas the *AD* of FCNN-2 represents the mean absolute deviation between the values of predicted *FOM*' and simulated *FOM*. In this study, we chose the *AD* rather than the mean relative error because the latter is strongly dependent on the magnitude of the *FOM* and therefore provides a less consistent assessment for the prediction accuracy. For example, for $W$ = 680 nm and $H$ = 280 nm, the simulated *FOM* is 1.523, and the predicted value of 1.554 results in a deviation of 0.027 and a relative error of ~ 2.0%. Whereas for $W$ = 340 nm and $H$ = 220 nm with a simulated *FOM* of 15.194, the same deviation of 0.027 results in a much smaller relative error of ~0.2%. A decrease in either $\Delta W$ or $\Delta H$ (*i.e.*, a larger training dataset) results in increased accuracy of FCNN-1 and decreased *AD* of FCNN-2, showing a trend consistent with the predicted *FOM*' distributions in Fig. 6. For comparison, the numbers of the ($W$, $H$) sets for training FCNN-1 and FCNN-2 are also shown in Fig. 7. At dataset No. 9, where the step sizes for both $\Delta W$ and $\Delta H$ are reduced to 20 nm, the numbers of training ($W$, $H$) sets increase to 396 for FCNN-1 and 281 for FCNN-2. Here, the accuracy of FCNN-1 exceeds 99.0%, and the *AD* of FCNN-2 decreases to < 0.05. This indicates that employing finer resolutions of the waveguide structural parameters enhances the capability of the FCNN framework to capture the dependence of the polarizer *FOM* on the waveguide geometry, thereby improving prediction accuracy. Given the high prediction accuracy achieved for both FCNN-1 and FCNN-2, we did not perform further analysis for the detailed error distributions in the small error ranges.

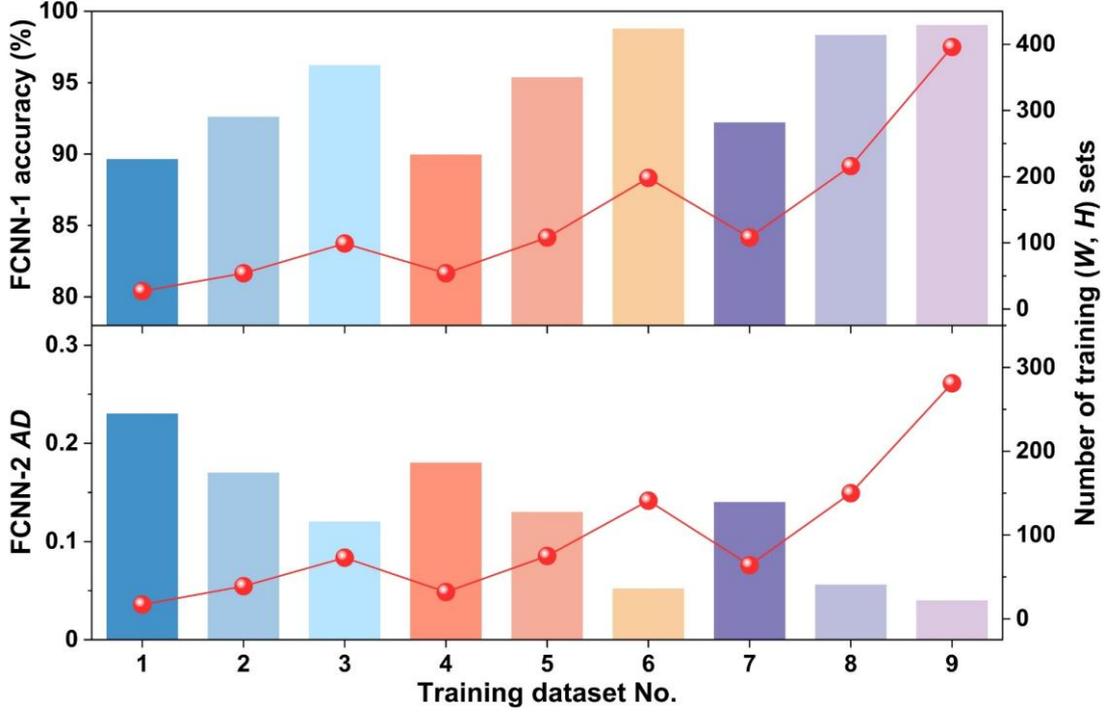

**Fig. 7.** Accuracy of FCNN-1 and average deviation (*AD*) of FCNN-2 for different training datasets (No. 1 – No. 9). The number of the (*W*, *H*) sets for each training dataset is also shown for comparison. The FCNN-1 for mode convergence identification and FCNN-2 for *FOM'* prediction were trained on the datasets listed in **Table 1**, and tested on the dataset with Δ' = 1 nm.

A more significant improvement in prediction accuracy is observed when Δ*W* or Δ*H* decreases from 80 to 40 nm, whereas the improvement becomes gradual when further decreasing the step size from 40 to 20 nm. For example, when Δ*W* remains fixed at 20 nm, progressively reducing Δ*H* from 80 nm to 40 nm and 20 nm results in two notable trends. In FCNN-1, the number of (*W*, *H*) samples in training datasets No. 4–6 increases from 108 to 216 and 396, accompanied by an accuracy improvement from 92.2% to 98.3% and 99.0%. In FCNN-2, the corresponding training sizes increase from 64 to 150 and 281, whereas the *AD* decreases from 0.14 to 0.06 and 0.04. This indicates that the accuracy of FCNN-1 and the *AD* of FCNN-2 do not follow a linear trend with the size of the training dataset. This trend arises because reducing the step size beyond a practical threshold adds additional (*W*, *H*) samples that do not reveal finer variations in the ($W$, $H$) – $FOM$ mapping, but instead produce redundant data with nearly identical performance characteristics. In addition, the performance of the FCNN framework shows greater sensitivity to $\Delta H$ than $\Delta W$, since small variations in $H$ lead to stronger changes in the effective refractive index and optical field distribution, resulting in larger differences in optical responses among samples. These findings establish the validity of the proposed FCNN framework within a theoretical and numerical context, and experimental validation of the predicted high-$FOM$ designs will be the subject of our future work.

Under a fixed training dataset condition with $\Delta H$ = 20 nm and $\Delta W$ = 20 nm, we also investigated how the configuration of the FCNN framework influences the prediction performance. Taking FCNN-1 as an example, a two-hidden-layer neural network configuration (32, 64) achieved a classification accuracy of 78.3%. The accuracy increased to ~99.0% for a three-hidden-layer configuration (32, 64, 32), and further to ~99.3% for a four-hidden-layer configuration (32, 64, 64, 32). When the number of the hidden layers was fixed at three, the configuration (16, 32, 16) achieved a classification accuracy of ~97.1%, whereas a larger-scale configuration (64, 128, 64) with more neurons in each layer yielded a slightly increased classification accuracy of ~99.1%. Considering the above testing results, a three-hidden-layer configuration (32, 64, 32) is chosen for FCNN-1 to balance the trade-off between improving accuracy and reducing training cost.

To quantitatively assess the computational advantage of the proposed FCNN approach, we compare its computing time with that of traditional mode simulations. In this work, we focus on designing 2D-material-based waveguide polarizers, and prior knowledge of optimal structural parameter regions is usually limited. In this context, exhaustive parameter scanning via traditional mode simulations represents a common workflow for broadly exploring the design space, and the full parameter scanning approach was adopted as the comparative baseline in our work. The total time required for mode-simulation-based method includes three parts: (i) conducting mode simulations via COMSOL Multiphysics, with the software running on a system with an Intel(R) Core (TM) i7-7700K CPU running at 4.20 GHz and 32.0 GB of RAM, (ii) manually identifying fundamental TE and TM modes and classifying convergence, and (iii) calculating the *FOM* values. By contrast, the computing time of the FCNN framework consists of two stages. One involves predicting mode convergence using only FCNN-1, and the other corresponds to predictions of *FOM'* using both FCNN-1 and FCNN-2. For comparison, all computational stages of our FCNN method were carried out on the same computer as that used for the mode simulation method. Under the same hardware conditions, the observed differences in computing time can therefore be attributed to the computational efficiency of the approaches.

For ten representative (*W'*, *H'*) parameter sets, conventional mode simulations require ~300–400 s per set on average, with more than 80% of the time consumed by the mode-solving process. By comparison, the trained FCNN framework completes mode convergence identification using FCNN-1 in ~0.01–0.03 s and predicts the *FOM'* using both FCNN-1 and FCNN-2 in only ~0.04–0.08 s, resulting in a speedup of more than 4 orders of magnitude relative to conventional simulation. Notably, the test dataset with Δ' = 1 nm contains 140,901 parameter sets. A brute-force sweep of all (*W'*, *H'*) combinations at 1 nm test step size would require mode simulation to consume approximately $10^7$ seconds, which corresponds to over 100 days of continuous operation, underscoring the prohibitive computational demands and costs. Conversely, our FCNN framework offers substantial savings in computing time and cost, requiring less than 40s to exhaustively sweep all (*W'*, *H'*) in a test dataset of the same size. This highlights the advantage of our FCNN framework in handling massive sets of structural parameters for polarizers' performance optimization. In addition, these results also reveal an inherent trade-off, where improvements in prediction accuracy need to be weighed against the cost related to building the training dataset.

Although the present study focuses on a fixed structural parameter setting, the operational principle of our FCNN framework is universal and can, in principle, be extended to the design of 2D-material-based optical polarizers under different conditions such as varying wavelength, non-fixed rGO optical constants, and alternative integrated material platforms. These aspects will be the subjects of our future work.

## 5. Conclusion

In summary, we propose and demonstrate a FCNN-based machine learning framework for optimizing the performance of rGO integrated waveguide polarizers. By using a small dataset of low-resolution mode simulation results to train the framework, it can rapidly and accurately predict the polarizer FOMs across a much larger structural parameter space with high resolution. The proposed framework achieves high prediction accuracy, with an *AD* below 0.05, whereas reducing overall computing time by more than 4 orders of magnitude compared with traditional mode simulation methods. Besides, nine training datasets with different sizes are constructed to analyze the trade-off between improvements in prediction accuracy and the cost of building the training dataset. These results demonstrate that the FCNN-based machine learning framework provides an efficient means for the design and optimization of rGO integrated waveguide polarizers.

## Funding

This work was supported by the Australian Research Council Centre of Excellence in Optical Microcombs for Breakthrough Science (Grant No. CE230100006), the Australian Research Council Discovery Projects Programs (Grant Nos. CE170100026, DP190103186, and FT210100806), Linkage Program (Grant Nos. LP210200345 and LP210100467), the Industrial Transformation Training Centers scheme (Grant No. IC180100005), the Beijing Natural Science Foundation (Grant No. Z180007), the National Natural Science Foundation of China (Grant No. 12404375), and the Innovation Program for Quantum Science and Technology (Grant No. 2021ZD0300703).

## Author contributions

**Rong Wang:** Writing – original draft, Investigation, Visualization, Formal analysis, Data curation, Validation, Software. **Yijun Wang:** Writing – Conceptualization, review & editing, Investigation, Validation. **Di Jin:** Writing – Conceptualization, review & editing, Investigation, Validation, Supervision. **Junkai Hu:** Investigation, Visualization, Formal analysis. **Wenbo Liu:** Investigation. **Yuning Zhang:** Investigation, Funding acquisition, Resources. **Baohua Jia:** Funding acquisition, Resources. **Duan Huang:** Funding acquisition, Supervision, Resources. **Jiayang Wu:** Conceptualization, Writing – review & editing, Supervision, Project administration. **David J. Moss:** Supervision, Funding acquisition, Project administration, Resources.

## Declaration of competing interest

The authors declare no competing interests.